%% file: main.tex
\documentclass[sigconf]{acmart}
\input{macros}

\copyrightyear{2026}
\acmYear{2026}
\setcopyright{cc}
\setcctype{by-nc-nd}
\acmConference[GoodIT '26]{International Conference on Information Technology for Social Good}{September 02--04, 2026}{Pisa, Italy}
\acmBooktitle{International Conference on Information Technology for Social Good (GoodIT '26), September 02--04, 2026, Pisa, Italy}
\acmDOI{10.1145/3794786.3830741}
\acmISBN{979-8-4007-2483-1/2026/09}

\graphicspath{{img/}}

\title{Watching What We Eat: Information Quality and Body Image in Diet-Related YouTube Videos}

\author{Maddalena Ghiotti}
\affiliation{%
  \institution{Politecnico di Torino, Turin, Italy}
  \institution{KTH Royal Institute of Technology, Stockholm, Sweden}
  \country{}
}
\author{Daniela Paolotti}
\affiliation{%
  \institution{ISI Foundation}
  \city{Turin}
  \country{Italy}
}
\author{Yelena Mejova}
\affiliation{%
  \institution{ISI Foundation}
  \city{Turin}
  \country{Italy}
}

\renewcommand{\shortauthors}{Ghiotti et al.}

\keywords{Information Quality; Eating Disorders; Diet; Body image; Social Media; Video Platforms; YouTube}

\begin{abstract}
The widespread use of social media—particularly image- and video-based platforms—has turned them into key sources of both normative and informational content related to health and diet.
This may contribute to the development of disordered eating behaviors or, potentially, eating disorders.
This study uses mixed-methods analysis applied to 3129 YouTube videos about diet and weight loss in order to quantify the level of risk of low-quality information and heightened focus on the body image. 
We adapt three quality measurement frameworks from the literature -- PRHISM, HONcode and SMEC -- to the online video context, perform manual annotation of a sample of the data, and design an LLM content characterization pipeline to score the videos on the quality of their content and the focus on the human body. 
Surprisingly, we find that videos around personal storytelling and mindset \& motivation are associated with higher-quality content, whereas supplement reviews (arguably more medically sensitive ones) are not. 
Further, the body-related mentions of weight measurement and negative body image are associated with an increased viewership, whereas the mentions of positive body image are associated with an increased engagement rate in terms of likes and comments, but not viewership. 
Worryingly, we find a cluster of videos categorized as ``music'' which promote the dietary supplements Mitolyn and the injectable weight loss drug Mounjaro. 
As video-based platforms grow in popularity, particularly among younger audiences, studies such as the one presented here are essential for developing empirically grounded tools to enhance the detection of harmful content and inform more effective moderation practices.

\end{abstract}

\ccsdesc[500]{Information systems~Social networks}
\ccsdesc[500]{Applied computing~Consumer health}
\ccsdesc[100]{Computing methodologies~Natural language processing}

\begin{document}

\maketitle

%%%%%%%%%%%%%%%%%%%%%%%%%%%%%%%%%%%%%%%%%%%%%%%%%%%%%%%%%%%%%%%%%%%%%%
\section{Introduction}
\label{sec:intro}
%%%%%%%%%%%%%%%%%%%%%%%%%%%%%%%%%%%%%%%%%%%%%%%%%%%%%%%%%%%%%%%%%%%%%%

Social media platforms have become central for information consumption and entertainment, particularly among younger people, with video-based platforms such as TikTok, YouTube, and Instagram experiencing rapid growth in both usage and influence. 
These environments increasingly shape how users learn about body norms, health behaviors, and lifestyle practices. 
Visual and short-form content, often delivered through algorithmically curated feeds, amplifies exposure to idealized body images and normative narratives around health and attractiveness. 
Diet-related content represents a particularly prominent and influential category within social media ecosystems, yet content analyses consistently reveal a lack of scientific rigor in online content, with popular videos often promoting rapid weight-loss strategies, restrictive diets, or anecdotal advice without expert validation \cite{basch_exploratory_2017, tang2022loss} and prioritizing aesthetic appeal over informational accuracy  \cite{zeng_whatieatinaday_2025}. 

Beyond issues of quality, there is substantial evidence linking exposure to such content with negative health outcomes. 
Studies have documented associations between diet-related social media use and increased body dissatisfaction, disordered eating behaviors, and harmful coping strategies \cite{sanzari2023impact}. 
Content genres such as \textit{fitspiration} and \textit{thinspiration} further exacerbate these risks by normalizing extreme body ideals and unhealthy practices \cite{tiggemann_exercise_2015, pearl_distinct_2016}. 
Emerging evidence suggests that younger users are especially susceptible to these dynamics, as they encounter repeated and often passive exposure to body-focused media through personalized recommendation systems \cite{mayoh_young_2021, davey2024risk}. 
These findings underscore the dual challenge posed by diet-related social media content: its widespread reach and its potential to misinform and harm users.
However, most current research focuses on the most popular content produced in a short period \cite{basch_exploratory_2017,tang2022loss,oksanen_pro-anorexia_2015} or that already being recommended by the algorithm \cite{zeng_whatieatinaday_2025}, instead of surveying the broader ecosystem of the diet-related content on these platforms. 

In this study, we aim to systematically assess the quality and potential risks of diet-related content on YouTube, the most popular video platform worldwide. 
To address the limitations of prior work, which has often relied on small-scale manual annotation and focused on popular content, we propose a scalable methodology combining automated transcript analysis with structured evaluation frameworks.
Specifically, we leverage large language models (LLMs) to annotate video transcripts according to the quality criteria and risk indicators developed and validated in the communications science literature. 
We formulate three research questions: 

\textbf{RQ1}: What is the information quality of diet-related content on YouTube and the extent of its focus on the body shape? 

\textbf{RQ2}: What video content attributes are associated with lower information quality and higher body focus? 

\textbf{RQ3}: To what extent does the quality and body focus relate to the user engagement with the content?

By integrating computational methods with validated assessment standards, this study contributes to a more comprehensive understanding of the informational landscape surrounding diet and health on social media.

%%%%%%%%%%%%%%%%%%%%%%%%%%%%%%%%%%%%%%%%%%%%%%%%%%%%%%%%%%%%%%%%%%%%%%
\vspace{-0.2cm}
\section{Related Work}
\label{sec:relwork}
%%%%%%%%%%%%%%%%%%%%%%%%%%%%%%%%%%%%%%%%%%%%%%%%%%%%%%%%%%%%%%%%%%%%%%

The evaluation of information quality has been a longstanding concern across multiple domains, leading to the development of standardized frameworks, survey instruments, and annotation methodologies. 
In the context of health information, tools such as the Health On the Net (HONcode) \cite{boyer1998health} and the more recent Principles for Health-related Information on Social Media (PRHISM) \cite{denniss_development_2022} provide structured criteria for assessing credibility, transparency, and evidence-based communication.
These frameworks have been applied in empirical studies to evaluate online content, revealing widespread deficiencies in quality and adherence to best practices \cite{denniss_fail_2024}. 
Surveys have also played a key role in understanding how users seek and evaluate information, with large-scale studies demonstrating high reliance on digital sources and limited verification with healthcare professionals \cite{fassier2016seeking}. 
Complementary research highlights users' difficulties in assessing credibility, particularly in environments characterized by information overload and heterogeneous sources \cite{mayoh_young_2021,kreft_use_2023,fassier2016seeking}. 
Instead of surveying content consumers about their perceptions, content analysis methods often employ detailed coding schemes to operationalize relevant dimensions such as accuracy, completeness, and source authority \cite{squires_informing_2023,charnock_discern_1998,silberg1997assessing,guardiola-wanden-berghe_evaluating_2011}. 
While these approaches provide nuanced insights, they are typically labor-intensive and difficult to scale, motivating the exploration of automated techniques that can extend quality assessment to larger and more diverse datasets. 

A substantial body of work has focused specifically on evaluating the quality of health and diet-related content on social media platforms, with increasing attention to video-based media. 
Early studies of YouTube identified significant gaps in the accuracy and reliability of nutrition-related videos, 
particularly among those with high view counts \cite{basch_exploratory_2017}. 
Subsequent research has expanded this analysis to other platforms, revealing consistent patterns of low-quality information and a lack of expert involvement. 
For instance, analyses of TikTok content demonstrate that highly engaging diet-related videos often prioritize visual appeal and personal narratives over evidence-based guidance \cite{zeng_whatieatinaday_2025}. 
Yet other content may promote conditions acknowledged to be dangerous to health, such as that circulating within pro-anorexia communities \cite{oksanen_pro-anorexia_2015}.
These studies frequently rely on manual annotation using predefined criteria, although some have begun to incorporate semi-automated approaches to handle larger datasets. 
In parallel, research has examined the potential harms associated with exposure to such content, linking it to body dissatisfaction, negative affect, and disordered eating behaviors \cite{jeronimo2022effects}. 
The convergence of findings across platforms and methodologies underscores the systemic nature of the problem. 
However, the quality assessment of these platforms has focused on the most popular content, neglecting the ``long tail'' of the posted videos that are available to the consumer. 
Further, the studies alternate between small samples of popular content examined manually (e.g. \cite{zeng_whatieatinaday_2025}) and shallow larger samples on which sentiment analysis tools are run (\cite{oksanen_pro-anorexia_2015}). 
In this work, we use a hybrid data collection approach to sample both longitudinal posting activity around the topic of dieting, as well as to capture the most popular actors, and use the latest LLM technology to deeply examine their quality.  

Recent advances in large language models (LLMs) have opened new possibilities for the automated annotation and analysis of social media data. 
Unlike traditional machine learning approaches, which require task-specific training data and often struggle with contextual nuance, LLMs demonstrate strong performance in zero-shot and few-shot settings. 
The prompting strategies employ the decomposition of tasks into interpretable sub-dimensions, use of rubric-based scoring, and iterative refinement to improve consistency and reliability \cite{la_rocca_multimodal_2025,khalil_evaluating_2025,nori_capabilities_2023,wang_are_2023}.
For example, models such as GPT-4 have been successfully applied to the evaluation of medical content, showing high agreement with expert assessments \cite{khalil_evaluating_2025,kalyan_survey_2023}. 
These capabilities enable the adaptation of established evaluation frameworks into prompt-based annotation schemes, allowing LLMs to assess multiple dimensions of content quality and risk simultaneously. 
In the context of social media, this is particularly valuable for analyzing transcript data derived from videos, which can be lengthy, unstructured, and linguistically diverse \cite{chi_watchwithme_2025}. 
LLMs can capture implicit meanings, identify misleading claims, and evaluate tone and framing, outperforming conventional NLP pipelines that rely on surface-level features \cite{yang2025llm,rose2024using}. 
Moreover, their flexibility allows researchers to integrate multiple annotation tasks within a single workflow, improving efficiency and consistency. 
As a result, LLM-based approaches, which we also use in this work, represent a significant advancement over traditional methods, offering both scalability and depth in the analysis of complex, real-world data.

%%%%%%%%%%%%%%%%%%%%%%%%%%%%%%%%%%%%%%%%%%%%%%%%%%%%%%%%%%%%%%%%%%%%%%
\section{Data \& Methods}
\label{sec:data}
%%%%%%%%%%%%%%%%%%%%%%%%%%%%%%%%%%%%%%%%%%%%%%%%%%%%%%%%%%%%%%%%%%%%%%

\subsection{Data Collection}

We collect data using two distinct procedures to obtain samples of both the videos uploaded to the platform on a daily basis (which we refer to as the daily sample) and the videos that gained popularity during the study period (the popular sample).

\textbf{Daily sample}. 
Between January 8 and June 7, 2025, we perform a continuous collection via the YouTube Data API \cite{APIref} by running daily retrievals of 500 most recent diet and weight-loss videos uploaded the previous day. 
We combine search keywords (\emph{diet, dieting, diets, weight loss, weightloss, fasting, nutrition}) in OR logic, and follow the methodology of \cite{mejova2025vaccine} to filter the results by the US region and English language (interestingly, the API still returned videos posted by channels identifying from outside of the US, as we show in the Results section). 
Specifically, we filter the videos using several criteria: English as the default audio language, duration between one and sixty minutes, and at least one keyword match in both the title and transcript (or at least three transcript matches if the title contained none). 
For each selected video, we collect metadata at upload time and then track it periodically — daily for the first week, then weekly up to three months — capturing evolving engagement statistics and a sample of comments.
Finally, because the Data API does not provide video transcripts, we extract them using the YouTube Transcript API \cite{youtubeTranscriptAPI}, which retrieves both manual and auto-generated captions. 
We automatically exclude any video lacking a transcript, map category IDs to human-readable names, drop records with malformed fields or missing engagement statistics.
This approach yielded 2870 videos with complete metadata. 

\textbf{Engagement}.
The engagement metrics associated with the collected videos include the number of views, likes, and comments.
These statistics are captured at the time of the video's collection. 
Thus, we consider the latest most complete available statistic we find available during the periodic re-collection, which is 70 days from upload for the daily sample.
Further, we compute the engagement rate, which is the sum of the number of likes and comments, normalized by the number of views.

\textbf{Popular sample}. 
To ensure that highly popular videos were not missed by the above daily search, we conduct a supplementary retrospective collection between November 10 and December 1, 2025. 
We target the same five-month window and query the API, but order results by view count across sliding time windows of one day, ten days, and one month, in batches of 50 (the API maximum per request). 
We then take the union of the three results. 
After the filtering steps as above, this popular sample yielded 384 videos. 
Combined with the daily sample, the final dataset consists of \num{3129} unique videos.

\vspace{-0.2cm}
\subsection{Content Characterization}

We summarize the topical characteristics of the sampled videos using three resources. 
First, we examine the list of topics YouTube API provides for each video. 
Second, we use Non-negative Matrix Factorization (NMF) to extract a set of topics from the video title, description, and transcript. 
Following \citet{cheng_multi-scale_2022}, we pre-process the text by removing URLs, numbers, and emojis, converting to lowercase, removing timestamps and text enclosed in square brackets, and other stop words. 
We use a combination of the topic coherence measure and manual selection of the topics to select the final number of topics, $k = 13$ (with coherence at $0.642$). 
Finally, we assign meaningful titles to the topics, for further analysis. 
A third source of information is that about the channel: the YouTube API provides a field describing the topical specialty of the posting channel.
We also manually classify the channel owner of a subset of videos (see the section below) following \cite{mejova2025vaccine} into institution, commercial, individual, or other, as well as into broad topical categories: doctor, health, science, news, opinion, lifestyle, or other.

Further, we compute a range of surface-level statistical features from the title, description, and transcript text to quantitatively characterize video content. 
These include basic length-based metrics such as title length (in characters), description length, and average word length in the transcript, which capture verbosity and linguistic complexity. 
Additional features quantify stylistic elements, such as the ratio of uppercase letters, frequency of exclamation marks, and presence of emojis, reflecting emphasis or sensationalism in presentation. 
The analysis also considers structural indicators such as the number of hashtags, mentions, and external links in the description, which may signal engagement strategies or informational richness. 
Together, these surface statistics provide a standardized, non-semantic representation of textual characteristics. 
We explore the association between these, and the topical variables obtained above, with the quality and body scores, described below.

\subsection{Annotation for Information Quality and Body-related Content}

Guided by the previous literature, we utilize previously validated questionnaires in order to measure the quality of the information in each video and the extent to which the videos refer to the human body. 
Specifically, for the former, we adopt relevant measures from the Principles for Health-related Information on Social Media (PRHISM) questionnaire \cite{denniss_development_2022}, the Health on the Net Foundation Code of Conduct (HONcode) \cite{boyer1998health}, and the Social Media Evaluation Checklist (SMEC) \cite{squires_informing_2023}.
The final list of dimensions is shown in Table \ref{tab:quality_principles}. 
We score each principle on a scale from 0 to 4, or mark it as not applicable, if video lacks enough information for rating. 
The overall quality score of a video is an average of the non-empty scores, rescaled to a range of 0 to 100 (with a higher score meaning higher quality). 

\begin{table*}[th] 
\caption{Information Quality Principles. Each principle includes citation to the questionnaire from which it was adapted: PRHISM \cite{denniss_development_2022}, HONcode \cite{boyer1998health}, and SMEC \cite{squires_informing_2023}.
\label{tab:quality_principles}}
\begin{center}
\renewcommand{\arraystretch}{1.45}
\footnotesize
\begin{tabular}{p{0.15\linewidth} p{0.80\linewidth}}
\toprule
Principle & Description \\
\midrule
    Authorship (PRHISM) & The speaker clearly presents her/his identity and relevant qualifications or experience (e.g., nutritionist, personal trainer, medical professional, or personal experience). \\
    \Ts Authoritative (PRHISM) & Information is within the speaker’s scope of expertise. \\
    \Ts Action-oriented (PRHISM) & Information is presented clearly and concisely, providing enough context to help viewers make informed decisions, while acknowledging that multiple approaches may exist and explaining possible alternatives. \\
    \Ts Attribution (PRHISM) & Information includes citations and hyperlinks (either verbally, in graphics, or in the video description) to the original source of information.\\
    \Ts Balance and justifiability (PRHISM) & Information is balanced and unbiased, avoiding exaggeration of benefits, minimization of risks, or promotion of ‘miracle’ solutions. \\
    \Ts Risks and benefits\newline (PRHISM) & Information clearly outlines the potential benefits and any associated risks or side effects of the products, behaviors, or practices being discussed.\\
    \Ts Complementary\newline information (PRHISM) & Information is complementary and not designed to replace the relationship between individuals and health professionals. Information includes statements encouraging individuals to discuss choices with a relevant health professional. \\
    \Ts Referrals and\newline support (PRHISM) & The speaker directs viewers to additional reliable resources, official guidelines, or support services when relevant (e.g., eating disorder help lines).\\
    \Ts Readability and\newline comprehensibility (PRHISM) & Information is presented in plain, everyday language, avoiding jargon, technical terms, abbreviations, or uncommon words, or providing clear explanations when such terms are necessary.\\
    \Ts Acknowledgment of\newline uncertainty (HONcode) & Video openly acknowledges the limitations of the evidence it presents and the fact that knowledge may evolve over time.\\
    \Ts Separation of interests (HONcode and SMEC) & Information is separate from financial, political, or ideological messages, including avoiding promotion of stereotypical attitudes and discriminatory actions about clinical or social issues.\\
    \Ts Data (HONcode) & Creator shares data to support statements.\\
\bottomrule
\end{tabular}
\end{center}

\vspace{0.5cm}
\caption{Select body-related aspects adapted from \citet{munro_diet_2024}.  
\label{tab:body_principles}}
\begin{center}
\footnotesize
\renewcommand{\arraystretch}{1.45}
\begin{tabular}{p{0.15\linewidth} p{0.80\linewidth}}
\toprule
Principle & Description \\
\midrule
    Weight\newline measurement & Content includes references to body-weight measurement, whether as absolute values or as mentions of weight lost or gained. (ex. ‘What I eat to lose 16 kg’.)  \\
    \Ts Mention of calories & Content includes mentions of the number of calories. (ex. ‘This is one pint of ice cream, 1000 calories. And this is 8 cups of protein ice cream, 385 calories’.)\\
    \Ts Referencing\newline body image & Directly referencing body (ex. Man filming his shirtless body in the mirror at different timepoints. ‘This is my body one day of dieting. One week of dieting. Two weeks of dieting’.)\\
    \Ts Negative portrayal\newline of body image & Content negatively refers to or negatively portrays body image. Body image is how we think and feel about ourselves physically (including our perceived sexual attractiveness) and how we believe others see us. (ex. A man starts working out to look more like the muscular men on TV his partner was admiring.)\\
    \Ts Positive portrayal\newline of body image & Content positively refers to or positively portrays body image. Body image is how we think and feel about ourselves physically (including our perceived sexual attractiveness) and how we believe others see us. (ex. ‘This outfit looks so good on me, I don’t care what you say, this dress looks so good on me’) \\
    \Ts Comparison  & Content includes comparisons of the body across at least two time points, eventually using body-checking techniques to illustrate changes in weight or appearance, for example, Day 1 of a diet plan vs. Day 20. (ex. ‘Diet journey. 83 kg59·6 kg’. Photos of a woman at different timepoints, with the weight at the time in text.) \\
\bottomrule
\end{tabular}
\end{center}
\end{table*}

To measure the extent to which the content refers to the human body, we use the body-related variables developed by \citet{munro_diet_2024}, listed in Table \ref{tab:body_principles}.
The variables are scored and aggregated similarly to those for information quality, resulting in a score ranging from 0 to 100 (with a higher score meaning more focus on the body).

We created a validation set by manually annotating 50 videos, with 25 randomly selected from the daily sample and 25 from the popular one. 
While one author annotated all 50 sampled videos, four external annotators labeled a subset of 20 videos. 
On these 20 videos, a consensus between the two annotators was reached for every score, and the scores for all videos were adjusted to achieve a consistent interpretation of the principles.
The annotation involved watching the full video, the duration of which ranged from 65 to 3194 sec, with an average of about 12 minutes.
During annotation, we also included a question assessing the extent of AI use in the video production (from 0 when no AI tools were visibly used, to 4 when entire video was AI-generated), and two questions designed to characterize the channel which posted the video adopted from \citet{mejova2025vaccine} (described in subsection above). 
Finally, we annotated for any specific brands that were mentioned in the video and validate the topics resulted from the NMF analysis. 

Next, we develop an automatic annotation pipeline using OpenAI's GPT-4.1 model to estimate each of the criteria in the two scores, given the video title, description, and transcript. 
We explore zero- and few-shot alternatives, prompting for the chain-of-thought explanations, as well as two temperature settings. 
Validated on the manually annotated subset, the performance of the different models is statistically indistinguishable, thus we use the simplest version: zero-shot prompting without the chain-of-thought explanations with temperature 0.1.
Compared to the manually annotated scores, it achieved a correlation of Spearman's $\rho = 0.669$ for the information quality and $\rho = 0.752$ for the body-related content.
We truncate long transcripts by retaining the concatenation of the first and last 1,500 words (3,000 words in total). 
This model is then used to annotate the entire dataset. 
In the aims of reproducibility, we make the data, annotations and code used for this study available online\footnote{\url{https://github.com/MaddalenaGhiotti/YouTubeWeightLossProject}} for the research community.

%%%%%%%%%%%%%%%%%%%%%%%%%%%%%%%%%%%%%%%%%%%%%%%%%%%%%%%%%%%%%%%%%%%%%%
\section{Results}
\label{sec:results}
%%%%%%%%%%%%%%%%%%%%%%%%%%%%%%%%%%%%%%%%%%%%%%%%%%%%%%%%%%%%%%%%%%%%%%

\subsection{Categories of Diet-related Content}

We begin by examining the sources of diet information we have captured in the samples. 
Concerning the categories of channel owner type and channel topical category identified by hand during annotation, we find that by far most accounts (58\%) are individual ones, and only 8\% are identifiably commercial, and only 1 account institutional (the rest were mostly removed channels which were difficult to characterize). 
Table \ref{tab:channelTopic} shows the distribution of most frequent channel categories (provided by the API, for the whole dataset), along with the number of channels and videos. 
As expected, \emph{Health}, \emph{Lifestyle}, and \emph{Physical fitness} top the list. 
\emph{Food} channels contribute proportionally most to the dataset (on average 1.93 videos per channel), however \emph{Health} and \emph{Lifestyle} also have relatively high posting rate (at around 1.74), indicating channels repeatedly returning to the topic.

\begin{table}[t] %htbp
\caption{Top channel topic categories (up to a large drop in videos), and their number of channels ($N_c$) and videos ($N_v$).
\label{tab:channelTopic}}
\begin{center}
\begin{tabular}{lrrr|lrrr}
    \toprule
     & \textbf{$N_c$} & \textbf{$N_v$}  & v/c &  & \textbf{$N_c$} & \textbf{$N_v$}  & v/c \\ 
    \midrule
    Health & 1569 & 2711 & 1.73 & Knowledge & 27 & 43 & 1.59 \\
    Lifestyle & 1170 & 2044 & 1.75 & Pet & 15 & 25 & 1.67 \\
    Phys. Fitness & 376 & 541 & 1.44 & Film & 12 & 12 & 1.00 \\
    Food & 297 & 572 & 1.93 & Hobby & 10 & 18 & 1.80 \\
    Society & 145 & 206 & 1.42 & Fashion & 9 & 13 & 1.44 \\
    Ent-ment & 104 & 130 & 1.25 & Politics & 8 & 9 & 1.13 \\
    TV program & 88 & 135 & 1.53 & &&&\\
    \bottomrule
\end{tabular}
\end{center}
\end{table}

\begin{figure}[t]
\centering
      \caption{Number of videos across the categories provided by YouTube (left) and the topics extracted using NMF (right). Note, a video can have only one YouTube category, but can have multiple NMF topics. }
      \label{fig:popChannelCategories}
      \includegraphics[width=\linewidth]{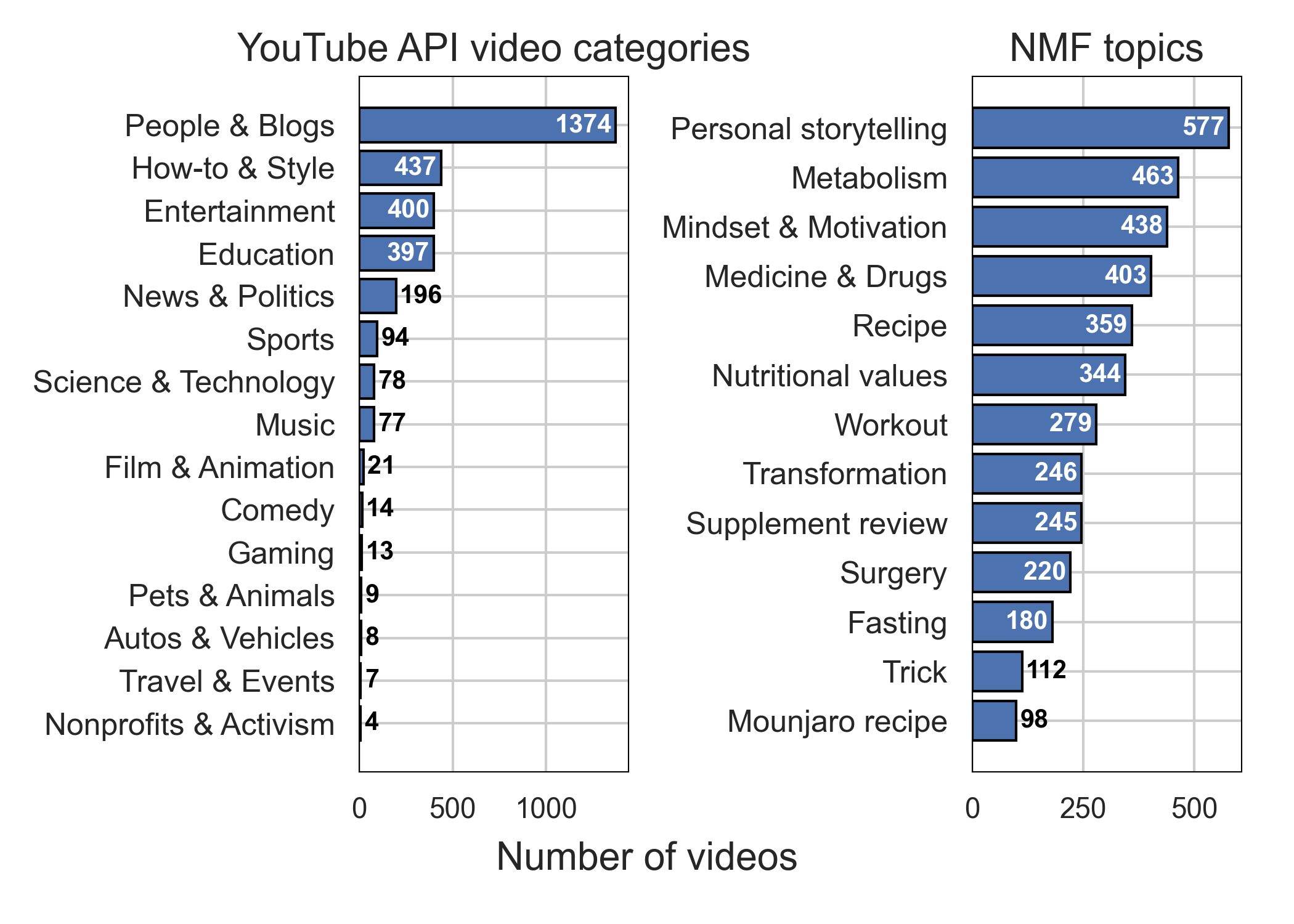} 
\end{figure}

When we compare the video categories provided by YouTube (Figure \ref{fig:popChannelCategories}, left) to those we extracted from the video content (right), we find that the platform ones are dominated by generic labels, mainly \emph{People \& Blogs}, % \textcolor{blue}{(??)} Y: reworded
followed by \emph{Entertainment} and \emph{Education}.
Instead, our NMF topics provide a fine-grained view of the data collection (note that a video may be assigned several NMF topics).
We find that \emph{personal storytelling} (disclosures of personal and emotional nature around the successes and challenges, mainly with weight loss) covers the most videos, followed by content focusing on the \emph{metabolism} (how calories are burned, blood sugar levels, and digestion), and on \emph{mindset \& motivation} (stress and emotion management, body satisfaction, weight loss habit maintenance). 
The smallest topic is also the most peculiar: it concerns the recipes meant to replicate the effects of the prescription drug Mounjaro (see the case study below). 
In the subsequent analyses of information quality and user engagement, we consider the NMF topics.

\subsection{Information Quality and Body-related Content}

\begin{figure}[t]
\vspace{-0.4cm}
\centering 
      \caption{Distribution of the quality and body scores, as well as the mean ($\mu$) and variance ($\sigma^2$). }
      \vspace{0.5cm}
      \includegraphics[width=\linewidth]{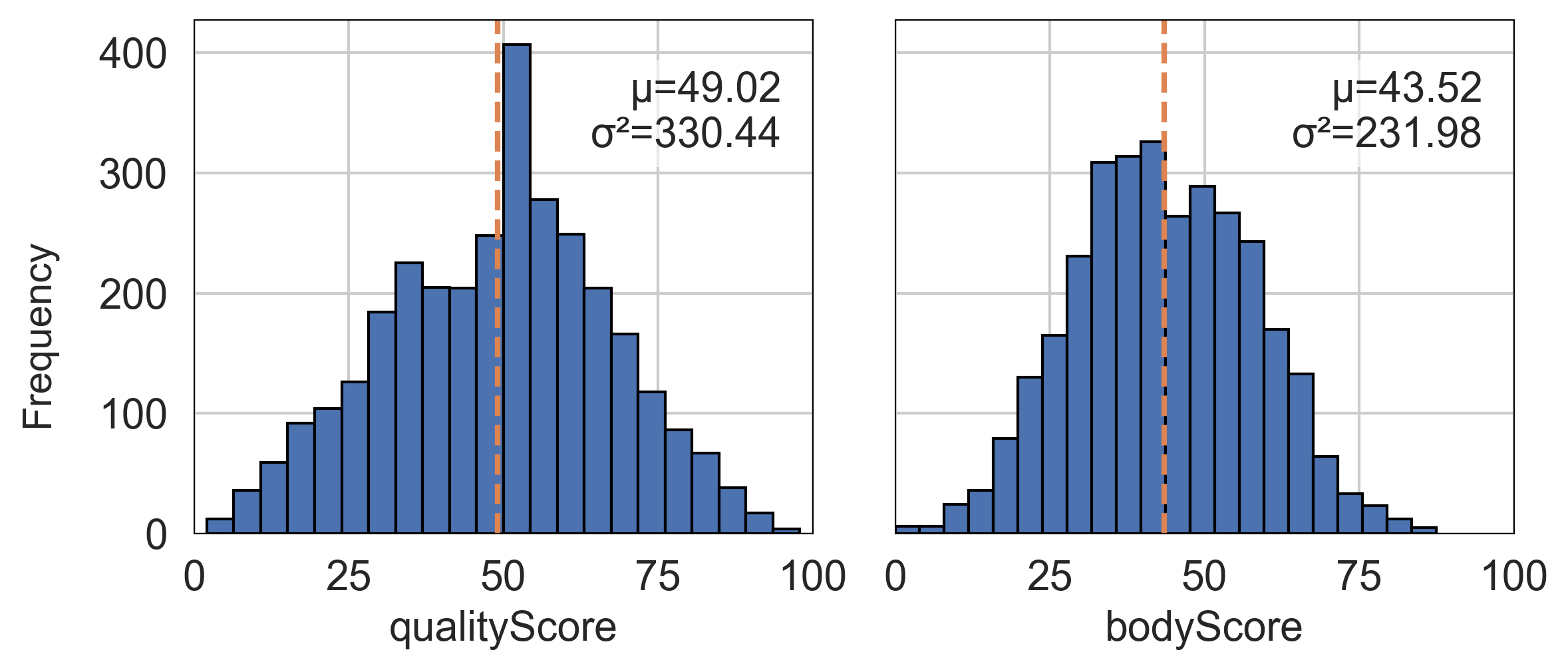}
      \label{fig:i_quality_body}
      \vspace{-0.7cm}
\end{figure}

Next, we turn to the estimates of the information quality (``quality score'') and the extent to which body-related content appears in the video (``body score''). 
Figure \ref{fig:i_quality_body} presents the distribution of the scores in the dataset. 
We find that few videos achieve high quality scores, with an average of around 49 out of 100. 
Similarly, the body score averages at 44 out of 100.
Note that the two scores correlated only slightly (Pearson $r=0.054$, $p=0.0024$), indicating that there are low-quality videos at both sides of the body score spectrum. 
Manual inspection revealed that the lowest-quality videos were those promoting ``easy'' diet ``tricks'' (``coffee loophole'', ``detox drinks'', and weight-loss challenges). 
Those scoring low on body score focus strictly on the food, those with high body scores often mention the body explicitly (``How To Starve Yourself to 8\% Body Fat'').

\begin{figure*}[t]
\caption{The distribution of the quality and body scores across (left) channel categories and (right) channel countries. \label{fig:q_boxplot_scores_across}}
\centering 
    \includegraphics[width=0.45\linewidth]{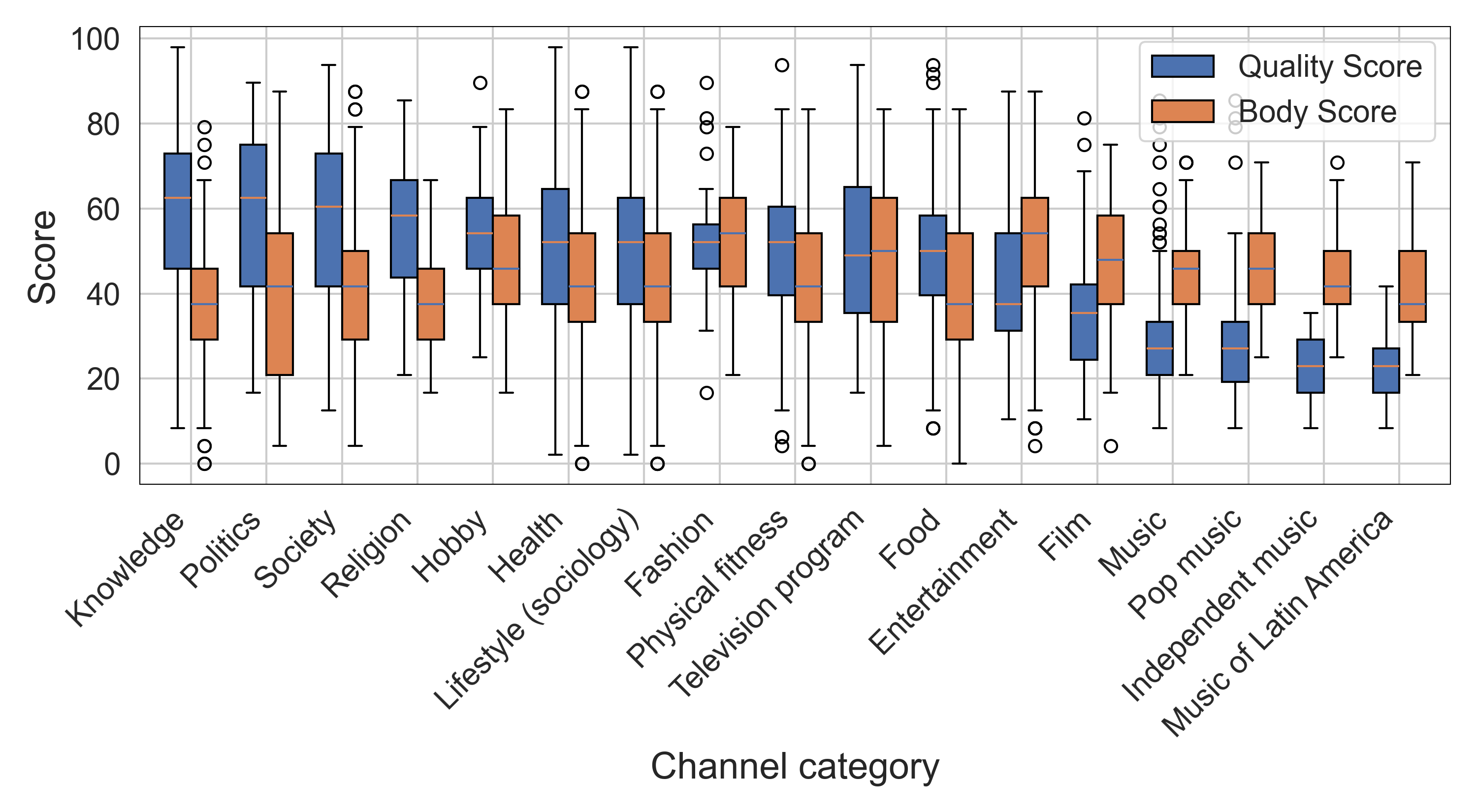}\hspace{0.5cm}
    \includegraphics[width=0.45\linewidth]{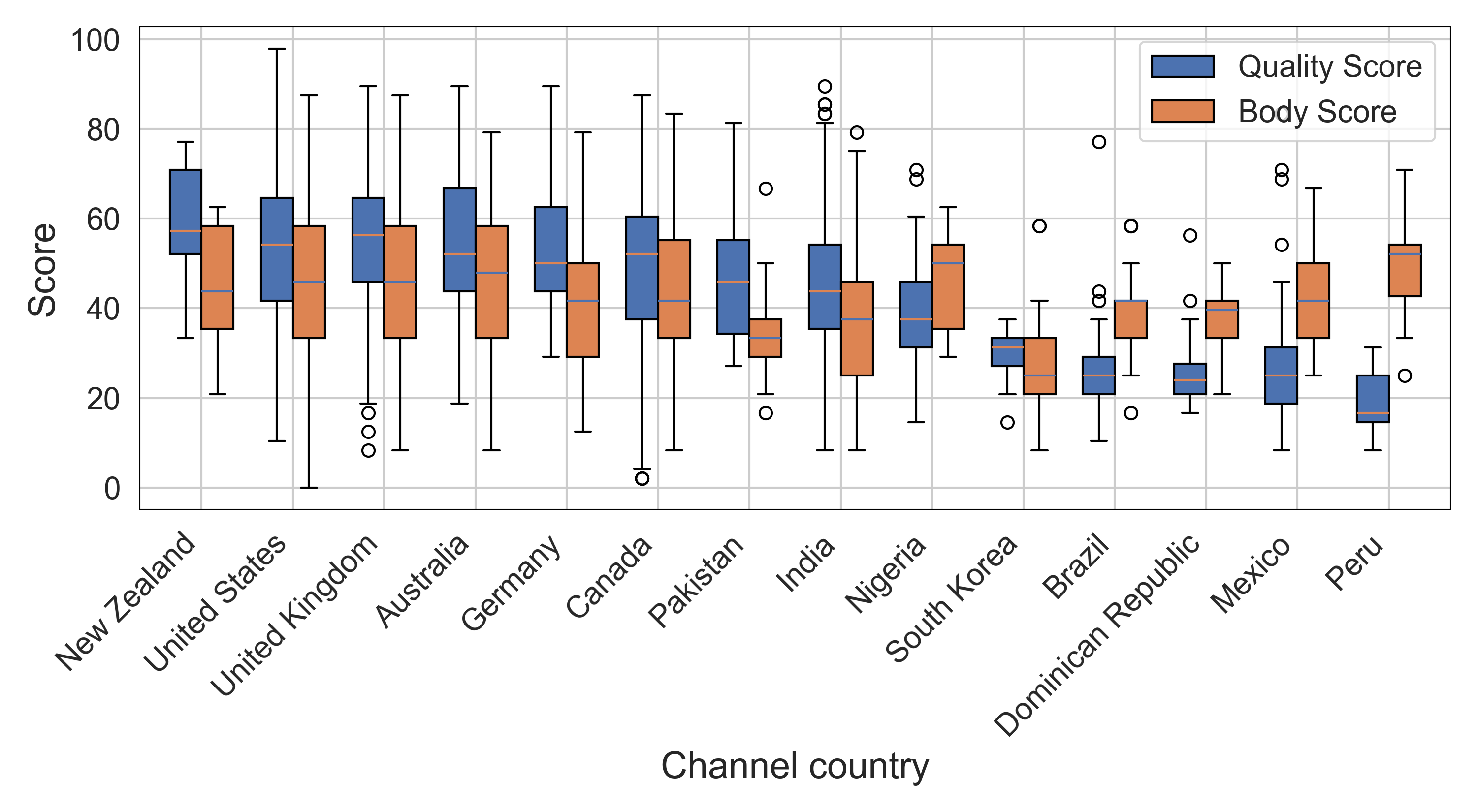}
\end{figure*}

\begin{table*}[t]
    \centering
    \caption{Spearman’s correlation coefficient ($\rho_s$) between the \textit{quality score} (left) or \textit{body score} (right) and descriptive variables of the video (or their logarithmic transformations) and the topic of the video. 
    Additionally, corresponding $p$-values ($p$) and Bonferroni-adjusted $p$-values ($p_{adj}$) intervals are presented. Within each of the two sections of the table, variables are reported in descending order of the absolute values of the coefficients. Confid. levels: $p_{adj} < 0.001$ ***, $p_{adj} < 0.01$ **, $p_{adj} < 0.05$ *.}
    \label{tab:quality_body_univCorr}
    \begin{tabular}{lrll | lrll}
    \toprule
        \textbf{Quality Score} &&&& \textbf{Body Score} &&&\\
         & \textbf{$\rho_s$} & \textbf{$p$} & \textbf{$p_{adj}$} & &  \textbf{$\rho_s$} & \textbf{$p$} & \textbf{$p_{adj}$} \\ 
        duration (secs) (log) & 0.413 & 3.034e-129 & *** &             avg word length (char) & -0.316 & 2.210e-73 & ***  \\ 
        \# mentions in description (log) & 0.248 & 3.381e-45 & ***  &  duration (secs) (log) & 0.311 & 2.707e-71 & ***  \\
        title length (char) & -0.241 & 1.711e-42 & ***  &              \# mentions in description (log) & 0.143 & 9.724e-16 & ***  \\
        avg word length (char) & -0.234 & 2.633e-40 & *** &                title\&desc uppercase ratio & 0.119 & 2.060e-11 & ***  \\ 
        channel description length & 0.224 & 8.116e-37 & ***  &        title\&desc exclamations ratio & 0.074 & 3.292e-05 & ***  \\ 
        \# links in description (log) & 0.152 & 1.485e-17 & ***  &     \# links in description (log) & 0.059 & 9.274e-04 & *  \\  
        channel video count (log) & -0.149 & 6.433e-17 & ***  &        \# hashtags in description (log) & -0.026 & 0.142 &   \\ 
        channel age (days) & 0.143 & 9.163e-16 & ***  &                description length (log) & 0.018 & 0.312 &   \\ 
        channel subscriber count (log) & -0.138 & 8.192e-15 & ***  &   title length (char) & 0.018 & 0.313 &   \\ 
        title\&desc uppercase ratio & -0.128 & 6.305e-13 & ***  &      channel video count (log) & -0.007 & 0.692 &  \\
        title\&desc emoji ratio & -0.123 & 4.856e-12 & ***  &          channel age (days) & -0.006 & 0.737 &  \\ 
        \# hashtags in description (log) & -0.045 & 0.011 &   &        channel description length & 0.005 & 0.763 &  \\ 
        title\&desc exclamations ratio & -0.037 & 0.041 &   &          channel subscriber count (log) & 0.004 & 0.826 &  \\ 
        description length (log) & -0.015 & 0.414 &   &                title\&desc emoji ratio & -9.320e-04 & 0.958 &  \\
        Medicine \& Drugs & 0.444 & 1.804e-151 & ***  &                Personal storytelling & 0.429 & 5.611e-140 & *** \Tf \\ 
        Mindset \& Motivation & 0.265 & 1.357e-51 & ***  &             Transformation & 0.212 & 3.372e-33 & ***  \\ 
        Personal storytelling & 0.246 & 1.591e-44 & ***  &             Metabolism & -0.211 & 9.073e-33 & ***  \\ 
        Surgery & 0.207 & 1.471e-31 & ***  &                           Mindset \& Motivation & 0.200 & 1.485e-29 & ***  \\ 
        Trick & -0.181 & 1.797e-24 & ***  &                            Workout & 0.142 & 1.529e-15 & *** \\
        Nutritional values & 0.163 & 3.664e-20 & ***  &                Nutritional values & 0.139 & 5.754e-15 & ***  \\ 
        Supplement review & -0.159 & 3.547e-19 & ***  &                Fasting & 0.137 & 1.231e-14 & ***  \\ 
        Mounjaro recipe & -0.145 & 3.617e-16 & ***  &                  Supplement review & -0.119 & 2.096e-11 & ***  \\ 
        Fasting & 0.091 & 3.484e-07 & ***  &                           Surgery & 0.118 & 2.969e-11 & ***  \\ 
        Workout & 0.087 & 1.031e-06 & ***  &                           Recipe & -0.085 & 1.791e-06 & ***  \\ 
        Recipe & -0.061 & 5.804e-04 & *  &                             Medicine \& Drugs & -0.046 & 0.010 &   \\ 
        Transformation & -0.049 & 0.007 &   &                          Trick & -0.035 & 0.049 &  \\
        Metabolism & -0.006 & 0.720 &   &                              Mounjaro recipe & 0.013 & 0.451 &   \\
        \bottomrule
    \end{tabular}
\end{table*}

Further, Figure \ref{fig:q_boxplot_scores_across} shows the distribution of the quality and body scores by the topic and the country of the posting channel. 
We find that the quality score is highest on average for channels in the categories of \emph{Knowledge}, \emph{Politics} and \emph{Society}, and the lowest in the categories of \emph{Music} (concerning which, see the case study in the section below). 
Although the data collection queries requested videos in the ``US region'', 31\% of channels of the returned videos identified themselves as another country (UK at 9.5\%, India 7.2\%, Canada 5.4\%, Australia 2.4\%, and most of the rest can be seen in the figure).
We find that the quality score is the highest for content from channels in New Zealand, the US and UK, Australia, Germany, and Canada. 
It is lowest for channels self-identifying as in Brazil, Dominican Republic, Mexico, and Peru (see Case Study subsection below on more regarding some Central and South American channels).

\begin{table*}[t]
    \centering
    \caption{Spearman’s correlation coefficient ($\rho_s$) between individual principle scores and the view count (left) and engagement rate (right), together with the corresponding $p$-values ($p$) and Bonferroni-adjusted $p$-values ($p_{adj}$) intervals. Below these, same is listed for significant correlations of extracted video topics. Scores are reported in descending order of the absolute values of the coefficients, and only results with $p_{adj} < 0.05$ are shown. Confidence levels: $p_{adj} < 0.001$ ***, $p_{adj} < 0.01$ **, $p_{adj} < 0.05$ *. 
    }
    \label{tab:engagement_correlation}
    \begin{tabular}{lrll|lrll}
	     \toprule
		  \textbf{Views} & & & & \textbf{Engagement Rate} & & & \\
         & $\rho_s$ & $p$ & $p_{adj}$ & & $\rho_s$ & $p$ & $p_{adj}$ \\ 
        Separation of interests & -0.202 & 5.618e-26 & *** &  Weight measurement & 0.149 & 1.915e-14 & ***  \\
        Negative B.I. & 0.151 & 4.670e-15 & *** & Complementary information & 0.133 & 7.513e-12 & ***  \\
        Weight measurement & 0.139 & 7.242e-13 & ***  & Authorship & 0.124 & 2.252e-10 & ***  \\ 
        Comparison & 0.128 & 3.606e-11 & ***  & Readability and comprehens. & 0.117 & 2.217e-09 & ***  \\
        Referencing B.I. & 0.137 & 1.535e-12 & ***  & Positive B.I. & 0.114 & 4.455e-09 & ***  \\ 
        Balance and justifiability & -0.115 & 3.033e-09 & ***  & Referrals and support & 0.087 & 9.268e-06 & ***  \\ 
        Risks and benefits & -0.100 & 2.154e-07 & ***  & Action-oriented & 0.082 & 2.813e-05 & ***  \\ 
        Authorship & 0.068 & 4.765e-04 & ** & Data & -0.079 & 5.390e-05 & ***  \\ 
        Mention of calories & 0.065 & 8.748e-04 & * &  Comparison & 0.066 & 6.916e-04 & *  \\
		  & & & & Mention of calories & 0.060 & 0.002 & *  \\ 
        Personal storytelling & 0.208 & 2.491e-27 & *** & Personal storytelling & 0.319 & 7.751e-63 & *** \Tf \\
        Mindset \& Motivation & -0.207 & 5.187e-27 & *** & Metabolism & -0.126 & 1.156e-10 & *** \\
        Metabolism & -0.141 & 2.825e-13 & *** & Medicine \& Drugs & -0.118 & 1.551e-09 & *** \\
        Recipe & 0.131 & 1.030e-11 & *** & Transformation & -0.116 & 3.052e-09 & *** \\
        Mounjaro recipe & 0.123 & 2.258e-10 & *** & Nutritional values & 0.112 & 9.305e-09 & *** \\
        Nutritional values & 0.109 & 1.885e-08 & *** & Recipe & 0.078 & 7.003e-05 & ** \\
        Supplement review & 0.072 & 1.927e-04 & ** & Workout & 0.074 & 1.586e-04 & ** \\
        & & & & Fasting & 0.061 & 0.002 & *  \\
		  \bottomrule
    \end{tabular}
\end{table*}

To understand what characteristics of the videos are most correlated with the quality and body scores, we perform univariate correlation analyses, shown in \autoref{tab:quality_body_univCorr} (because of the complex relationships between the variables, we report the univariate analysis instead of the multivariate regressions, which require exclusion of variables due to multicollinearity). 
We log-transform highly skewed variables,
and use Spearman’s rank correlation coefficient ($\rho_s$), which is a non-parametric measure of the strength and direction of a monotonic relationship between two variables which may be skewed or non-normal.

Concerning the quality score (left half of the table), the largest positive effect size is observed for the video duration, followed by the number of mentions in the description. 
The lengths of content provide a contradictory picture: the length of the title and the average length of the words used in the text (title, description, and transcript) are negatively correlated with the quality score, while length of the channel description -- positively.
This may point to the practice of video title (but not description) ``stuffing'' with content for maximizing retrieval or recommendation \cite{zuze2013keyword}.
Surprisingly, the number of subscribers a channel has is negatively related to the quality score, pointing to the popularity of those posting such content. 

Considering the topic-related variables, several statistically significant associations are also observed. The strongest positive effect size is associated with \textit{Medicine \& Drugs}, followed by \textit{Mindset \& Motivation}. This result is consistent with expectations: the former topic is more closely related to medical content, which is more likely to be communicated by professionals or experts, while the latter typically involves motivational narratives rather than prescriptive claims or technical explanations, thereby reducing the likelihood of low quality.
Conversely, strong negative correlations are observed for \textit{Trick}, \textit{Supplement review}, and \textit{Mounjaro recipe}. These topics are more likely to involve non-expert creators presenting scientific claims or health-related advice, increasing the risk of lower-quality information. This is aligned with the aforementioned manual inspection.

Note that during the manual annotation of the 50-video subsample, we find a significant correlation between the level of AI use and the \textit{quality score} (not shown in this table). 
The estimated coefficient is $\rho_s = -0.466$, with a $p$-value of approximately 0.001, indicating that higher levels of AI-assisted content generation are associated with lower video quality.

Turning to the body score (right half of the table), a smaller number of variables, particularly non-semantic ones, show statistically significant associations. 
We find that the videos using shorter words are more likely to have body-related aspects, as well as those with higher duration.  
The presence of uppercase and exclamation marks in the title and description are also associated with body-related mentions, indicating there are stylistic peculiarities to such content.
The strongest positive correlation is observed for the topic of \textit{Personal storytelling}.
It is followed by \textit{Transformation}, \textit{Workout}, and, somewhat unexpectedly, \textit{Mindset \& Motivation}.
Negative, although moderate, correlations are observed for topics such as \textit{Metabolism}, \textit{Recipe}, and \textit{Supplement review}. 
This suggests that, although videos aligned with the \textit{Supplement review} topic may exhibit lower informational quality in previous analyses, their primary focus is less explicitly centered on physical appearance and body-related content. 
Finally, in contrast to the findings for the quality score, the level of AI use in video production, manually annotated for the 50-video subsample does not exhibit a statistically significant correlation with the body score. 

Although no AI-detection tools were used in this study, we can report the assessment of the annotators, who reported 11 
out of 50 videos (22\%) were categorized by the labelers as having AI tools were used somewhere in the video’s production. Out of these, in 8 videos, AI tools were used to modify or generate a substantial portion of the video, significantly influencing its content, although the video was not entirely AI-generated.

\subsection{User Engagement}

Recall that we can measure engagement in two ways: the number of views the videos have received and the engagement rate, which is the sum of the number of likes and comments, normalized by the number of views.
Due to the difference in collection timings, we perform this analysis on the daily collection only.
We find that the correlation between the quality score and both metrics is low: $\rho_s = -0.053$ with the view count and $0.059$ with the engagement rate. 
The relationship is slightly stronger for the body score, at $0.157$ with the view count and $0.112$ with the engagement rate. 
Table \ref{tab:engagement_correlation} shows the significant correlations between individual principles (the principles that comprise the quality and body scores) and the number of views the video receives (left) or the engagement rate (right).
We also include the correlations with the presence of specific topics.

Nearly all of the principles comprising the body score, e.g. \emph{weight measurement}, \emph{comparison} and \emph{negative body image}, are mainly positively correlated with the view count, indicating a viewer preference to the content focused on the human body. 
We find a similar pattern with engagement rate, however there it is the \emph{positive body image} that is positively associated with the engagement rate. 
Unfortunately, we find several important signals of quality negatively related to views: the declaration of the \emph{separation of interests}, \emph{balance and justifiability} and the disclaimer of \emph{risks and benefits}. 
However, being clear about the video's \emph{authorship}, providing \emph{complementary information} and \emph{referrals}, and higher \emph{readability and comprehensibility}, as well as being \emph{action-oriented} are all principles associated with greater engagement rate. 

Considering the video topics, those around \emph{personal storytelling} and \emph{recipe} (including \emph{Mounjaro recipe}, see case study below) are positively associated with the views, whereas those mentioning \emph{mindset \& motivation} or the \emph{metabolism} are associated with fewer views.
The engagement rate is also highest for the \emph{personal storytelling} videos, however the mentions of \emph{metabolism}, \emph{medicines \& drugs} and \emph{transformation} are associated with a lower engagement rate. 
These results point to the theme of \emph{personal storytelling} as the most engaging to the platform's audience.

\subsection{Case Study: Mitolyn and Mounjaro}

During the topic discovery, we noted the appearance of two pharmaceuticals: Mitolyn, a dietary supplement (a ``proprietary blend of 6 exotic nutrients and plants'', according to the company's website \cite{mitolyn}) and Mounjaro (generic drug name: tirzepatide), a gastric inhibitory polypeptide analog and a GLP-1 receptor agonist that is used for managing type 2 diabetes \cite{mounjaro}.
The videos mentioning these products occur unusually often in content posted by channels self-categorizing as ``music''. 
Among the 28 channels that uploaded videos associated with Mitolyn, the majority came from Central or South America, with Brazil (7 channels) and Mexico (6 channels) dominating the ranking. They are followed by the Dominican Republic, Peru, Chile, Ecuador, Venezuela, and Colombia. 

\begin{figure}[t]
\centering 
     %\vspace{-0.3cm}
      \caption{Example of a channel that first posted music videos, and later drug reviews.}
      \label{fig:mitolyn_example}
      \includegraphics[width=\linewidth]{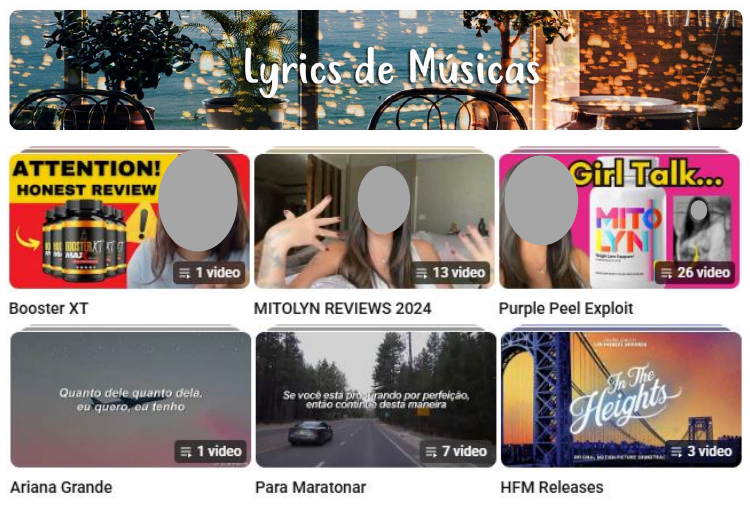}
\end{figure}

Upon manual inspection, we found that several of these channels were originally created to post music content (and some still appear to do so at the time of writing), but were later repurposed to share videos promoting pharmaceuticals. Notably, the same three women appear in multiple videos across different channels, suggesting the presence of a single producer behind this content.
An example of such a channel can be seen in Figure \ref{fig:mitolyn_example}.
At the time of writing, the medical ``review'' videos on this channel regularly garner over 1K views. 

%%%%%%%%%%%%%%%%%%%%%%%%%%%%%%%%%%%%%%%%%%%%%%%%%%%%%%%%%%%%%%%%%%%%%%
\section{Discussion \& Conclusions}
\label{sec:discussion}
%%%%%%%%%%%%%%%%%%%%%%%%%%%%%%%%%%%%%%%%%%%%%%%%%%%%%%%%%%%%%%%%%%%%%%

This study offers two key contributions to advancing the understanding of the quality of health-related content shared on social media. 
First, we offer a sample of diet-related YouTube videos that combine the daily activity of the platform's posters with the most popular content over the same time period, capturing both posters' efforts in this domain and the most successful ones. 
Second, we adapt existing quality assessment rubrics from the literature to this specific domain, and, using these, develop and evaluate an LLM-based approach for applying them at scale to thousands of videos.

We show that our dataset displays a wide range of quality scores, with an average of about 49 out of 100, indicating much room for improvement. 
Interestingly, our results indicate that information quality and the degree of body emphasis represent independent (largely uncorrelated) dimensions within diet-related discourse, and each will require targeted attention from researchers.
For instance, whereas we find that the mentions of weight measurement and negative body image are associated with the view count of the videos, the mentions of the positive body image are associated with an increased engagement in terms of likes and comments per view. 
The fact that the consumption of negative body-related messaging is more prominent than that of positive one, is especially worrying, since passive social media use has been shown to be related to body dissatisfaction and eating concerns (especially in young women) \cite{XIANG2024107360}, while the consumption of body positivity content may not have protective effects \cite{sanzari2023impact}.

Although dieting involves both medical and practical dimensions, the dominant theme in our dataset is personal storytelling—reflected not only in the volume of videos posted, but also in its strong positive link with views and the engagement rate.
This illustrates how personal accounts (including highly popular “influencer” profiles) can undermine or complicate the official messages that health authorities share on the platform.
Worryingly, we find that a channel's number of subscribers is negatively related to the quality score of its diet-related content.
Efforts are already underway to offer influencers training and support so they can produce higher-quality health-related content \cite{glauser2026influencing}; however, additional work is required to evaluate the reach and impact of these initiatives.

An especially egregious example of platform manipulation can be seen in our case study of Mitolyn/Mounjaro promotion by accounts first established as music channels. 
Such subscriber ``bait-and-switch'' has been found on, for instance, Twitter, with the ``repurposed'' accounts often used to promote political misinformation \cite{elmas2023misleading}.
A more extensive review is needed to determine whether some countries are more prone to hosting this type of content—knowledge that is important for both YouTube’s quality assurance team and public health authorities.
Note that these videos were identified using NMF topic discovery, a relatively lower-cost solution compared to LLM-based annotation.
Further, the manual annotation effort in this work has revealed a worrying trend: the labelers detected AI use in 22\% of the annotated videos, 
and the level of AI use was negatively related to the quality score, indicating its association with problematic content. 
Nonetheless, the role of AI in health-related communication will inevitably expand, and its impact needs to be carefully tracked by the research community.

This study has several limitations. 
Because of resource constraints, data collection was restricted to the U.S. Although some videos from other countries were included (due to the API’s definition of ``region''), the sample primarily reflects content seen by U.S. users and likely does not generalize to other countries. 
Targeted audits in other Central and South American countries are needed to assess the quality of their diet-related content. 
The LLM-based scorer also underperforms human evaluators, introducing potential error. 
Although we apply Bonferroni correction for multiple hypothesis testing, many observed correlations are small, suggesting no single strong determinant of diet-related content quality. 
In the future, including the visual information (alongside the voice transcripts) would improve the detection of body-related focus.
We encourage further development of rubrics and metrics for evaluating and auditing health information on multimedia platforms.

\bibliographystyle{ACM-Reference-Format}
\bibliography{references}

\end{document}

%% file: macros.tex
\usepackage{type1cm} % type1 computer modern font
\usepackage{graphicx} % advanced figures
\usepackage{xspace} % fix space in macros
\usepackage{balance} % to better equalize the last page
\usepackage{booktabs} % nicer tables
\usepackage{multirow} % multi rows for tables
\usepackage[font={bf}, tableposition=top]{caption} % captions on top for tables
\usepackage{bold-extra} % bold + {small capital, italic}
\usepackage{siunitx} % \num for decimal grouping
\usepackage[vlined,linesnumbered,ruled,noend]{algorithm2e} % algorithms
\usepackage{microtype} % compress text
\usepackage{xfrac} % nicer slanted fractions
\usepackage{mathtools} % amsmath++
\PassOptionsToPackage{hyphens}{url} % acmart compatibility
\PassOptionsToPackage{bookmarks, pdftex, colorlinks=true, pagebackref=true, backref=page}{hyperref} % acmart compatibility
\usepackage{cleveref} % smart references
\PassOptionsToPackage{square,numbers}{natbib} % acmart compatibility
\usepackage[hyperpageref]{backref} % back references
\usepackage{hyphenat} % name in single line
\usepackage{subcaption} % subfloats
\usepackage[export]{adjustbox} % valign
\renewcommand*\backref[1]{\ifx#1\relax \else (Cited on #1) \fi}

\newcommand\Tf{\rule{0pt}{2.8ex}} 
\newcommand\Ts{\rule{0pt}{2.2ex}}